\documentclass[prl,twocolumn]{revtex4}

\usepackage[dvips]{graphicx}

\usepackage{amsfonts}
\usepackage{amssymb}
\usepackage{amsmath}
\usepackage{bbm}
\usepackage{graphicx}

\newcommand{\ket}[1]{\text{$|#1\rangle$}}
\newcommand{\bra}[1]{\text{$\langle#1|$}}
\newcommand{\vac}{\ket{\text{vac}}}
\newcommand{\sqv}{\text{$\ket{\text{vac}(\eta)}_{\text{sq}}$}}

\begin{document}

\title{Geometric phase induced by a cyclically evolving squeezed vacuum
reservoir}
\author{Angelo Carollo$^{1, 2}$, G. Massimo Palma$^{3, 4} $, Artur {\L}ozinski$^3$, Marcelo Fran\c{c}a Santos$^{5}$ and Vlatko Vedral$^{6}$}
\address{$^{1}$Centre for Quantum Computation, DAMTP, University of
Cambridge, Wilberforce Road, Cambridge CB3 0WA, UK\\$^{2}$Institute for Theoretical Physics, University of Innsbruck,Technikerstra{\ss}e 21a, A-6020 Innsbruck, Austria\\  $^3$NEST \& Dip. Tecnologie dell'Informazione, Universit\'{a} di Milano;
via Bramante 65, I-26013 Crema (CR), Italy \\ $^4$ Dip. di Scienze Fisiche ed Astronomiche, Universit\'a di Palermo, via
Archirafi 36, I-90123 Palermo, Italy\\
$^5$University of Minas Gerais, Belo Orizonte, Brazil\\
$^6$School of Physics \& Astronomy, University of Leeds, LS2 9JT, United Kingdom}

\begin{abstract}
We propose a new way to generate an observable geometric phase by means of a completely incoherent phenomenon. We show how to
imprint a geometric phase to a system by "adiabatically" manipulating the environment with which it interacts. As a specific
scheme we analyse a multilevel atom interacting with a broad-band squeezed vacuum bosonic bath. As the squeezing parameters are
smoothly changed in time along a closed loop, the ground state of the system acquires a geometric phase. We propose also a scheme
to measure such geometric phase by means of a suitable polarization detection.
\end{abstract}

\maketitle

Whenever a pure quantum state undergoes a parallel transport along a closed path, it gathers information on the geometric
structure of the Hilbert space in which it lies. In this letter we will show that a a possible way to generate such a parallel
transport is by way of an irreversible quantum evolution. In several models of interaction with the environment there are some
"protected" subspaces, like the decoherence free subspaces (DFS), which are left unaffected~\cite{PalmaSE96}. 
 States lying in these 
subspaces are \emph{stationary}, i.e. they do
not evolve in time. A typical example is the ground state of an atomic system, which, trivially, remain unaffected by the coupling
with the electromagnetic field. However, there are situations in which the interaction between a system and an engineered
environment can generate  non-trivial ground states~\cite{EkertPBK89,AgarwalP90,PoyatosCZ96,MyattKTSKIMW00,KuzmichMP97}. For
instance, when a group of atoms collectively interacts with a broad band squeezed vacuum, the highly non-classical correlations
which are present in the field are transferred to the atomic system, which relaxes in a complex pure equilibrium state. In such a
scenario, the control over the engineered reservoir allows an indirect control on the state of the system to which it is coupled~\cite{CarvalhoMMD01}. Of particular interest is the possibility to change in time the reservoir parameters in such a way that the
"protected" system subspace evolves in a controlled fashion. Here we show that if this change in time is made slowly enough,  a
state lying in such a subspace evolves coherently and acquires information about the geometry of the space explored.

As an explicit example, we consider a suitable multilevel atomic system interacting with a broad band squeezed vacuum. To be more
specific let us consider first a three level atom whose interaction with an electromagnetic field in the rotating wave
approximation is described by the following Hamiltonian:
\begin{equation}
         H = H_S +\int \hat{a}^{\dagger}(\omega)\hat{a}(\omega)d\omega
         + \int \left[ g(\omega)S^{\dagger}\hat{a}(\omega) + H.c.
         \right] d\omega \text{,}\nonumber
\end{equation}
where $H_S = \Omega\sum_{k=-1}^1k\ket{k}\bra{k}$ is the free atomic Hamiltonian, $S = \ket{-1}\bra{0} + \ket{0}\bra{1}$ is the
atomic operator describing the absorption of an excitation and $\hat{a}(\omega)$ is the annihilation operator of the mode with
frequency $\omega$  ($\hbar =1)$.  The field, which we treat as a reservoir, is assumed to be in a broad band squeezed vacuum
state. In mathematical terms, this is obtained from the ordinary field vacuum state by means of the unitary operator
$\hat{K}(\eta)$
\begin{equation}
         \sqv = \hat{K}(\eta)\vac \text{,}
\end{equation}
where
\begin{equation}\label{eq:squeezer}
         \hat{K}(\eta)= \exp \frac{1}{2}\left\{\int \left[\eta
\hat{a}^\dag(\Omega-\omega)\hat{a}^\dag(\Omega+\omega) - H.c.\right] d\omega\right\}
\end{equation}
is a multimode squeezing transformation~\cite{PalmaSE96,AgarwalP90}, which correlates symmetrical pairs of modes around the
carrier frequency $\Omega$ and $\eta = e^{i\varphi}r$ is the squeezing parameter, whose polar coordinates $\varphi\in\{0\dots
2\pi\}$ and $r>0$ are called phase and amplitude of the squeezing, respectively.

The use of the Born Markov approximation, justified by the broadband nature of the field, leads to the following master equation
for the atomic degrees of freedom~\cite{PalmaSE96,AgarwalP90}:
\begin{equation}
\label{MasterEq} \frac{d\rho}{dt} =-\frac{\Gamma}{2}\{R^\dag R
\rho+\rho R^\dag R-2 R \rho R^\dag\}
\end{equation}
where $\Gamma = 2\pi | g(\Omega )|^2$, and
\begin{equation}
\label{Rmatrix} R(\eta)=S \cosh r+e^{i\varphi}S^\dag \sinh r.
\end{equation}
From (\ref{Rmatrix}) follows that the state
\begin{equation}
\label{eq:DFstate} \ket{\psi_{DF}(\eta)}=c \ket{-1}-e^{i\varphi}s
\ket{1},
\end{equation}
with $ c(r) = \frac{ \cosh r}{\sqrt{ \cosh 2 r }} $ and $s(r)=
\frac{ \sinh r}{ \sqrt{ \cosh 2 r }}$, satisfies
$R(\eta)\ket{\psi_{DF}(\eta)}=0$. In other words, this state is
unaffected by the environment, i.e. it is decoherence free.
Moreover $\ket{\psi(\eta)}$ represents the new ground state, as
all the other states of the atomic system relax to it.

As anticipated, the key idea is to smoothly change the squeezing parameter of the field in order to "adiabatically" drag a state
initially prepared in $\ket{\psi_{DF}(\eta_0)}$ into $\ket{\psi_{DF}(\eta_t)}$, where $\eta_t$ is the time dependent squeezing
parameter. We will show the existence of an "adiabatic" limit such that the transition probability of $\ket{\psi_{DF}(\eta)}$ to
the orthogonal subspace vanishes as the rate of change of $\eta$ becomes sufficiently small. Furthermore, we will show that after
a cyclic evolution of  $\eta$, the state $\ket{\psi_{DF}}$ acquires a geometric phase. It is worth stressing that this procedure,
although reminiscent of the usual adiabatic evolution, is a different physical phenomenon. The usual adiabatic approximation
refers to a coherent evolution, obtained by tuning the parameters of the system Hamiltonian, while the "steering process"
discussed here is achieved manipulating the environment. The essential difference is that in the latter case the system state can
be adiabatically controlled entirely by means of an incoherent phenomenon and no Hamiltonian term contributes to its time
evolution. To show how this incoherent adiabatic steering process can take place, consider the time dependent version of
equation~(\ref{MasterEq}) where $R(\eta_t)$ is explicitly dependent on time through $\eta_t$. It is useful to express the equation
of motion in the reference frame where $\psi_{DF}$ is time independent. To this end, consider the following unitary transformation
\begin{equation}
\label{ } O(\eta)=\begin{pmatrix}
      c(r)e^{-i\frac{\varphi}{2}}&0&s(r) e^{i\frac{\varphi}{2}}\\
      0&1&0   \\
    -s(r)  e^{-i\frac{\varphi}{2}}&0&c(r) e^{i\frac{\varphi}{2}}\\
\end{pmatrix},
\end{equation}
from the basis $\ket{1}$, $\ket{0}$, $\ket{-1}$ to the time
dependent basis $\ket{\tilde{1}}$, $\ket{\tilde{0}}$,
$\ket{-\tilde{1}}$, where $\ket{\psi_{DF}}$ coincides with
$\ket{-\tilde{1}}$ Under this change of frame, the equation of
motion becomes
\begin{equation}\label{eq:newmastereq}
         \frac{d\tilde{\rho}}{dt} = -\frac{\tilde{\Gamma}}{2}\left(
\tilde{R}^{\dagger}\tilde{R}\tilde{\rho}
         +\tilde{\rho} \tilde{R}^{\dagger}\tilde{R} - 2\tilde{R}\tilde{\rho}
\tilde{R}^{\dagger} \right)
-i[G,\tilde{\rho}]\text{,}
\end{equation}
where, in this new frame, $\tilde{\rho}=O(\eta_t)\rho O^\dag (\eta_t)$, $\tilde{R}=O(\eta_t) R(\eta_t) O^\dag (\eta_t)/\sqrt{\cosh
2r}$, $\tilde{\Gamma}=\Gamma \cosh 2 r$ and $G=i dO/dt O^\dag$ is a Hamiltonian term arising from the change of picture. Moreover,
in this frame the Lindbladian term, $\tilde{R}^\dag \tilde{R}$, assumes a simple diagonal form:

\begin{equation}
\tilde{R}^\dag \tilde{R}=
\ket{\tilde{1}}\bra{\tilde{1}}+\ket{\tilde{0}}\bra{\tilde{0}}.
\end{equation}

The main advantage of this transformation is that it allows to formulate clearly the adiabatic condition, since the rate of change
of the environment parameters are contained in the operator $G$. The limit that we are interested in is the one in which the
dominant contribution in equation~(\ref{eq:newmastereq}) comes from the incoherent terms, i.e. $|G|\ll \tilde{\Gamma}$.

An interesting case is the one in which the squeezing amplitude
 is kept constant while its phase is slowly changed  from
$0$ to $2\pi$. This adiabatic evolution can be easily achieved by tuning, for example, the carrier frequency, $2\Omega$, of the squeezed state slightly off resonance from the two photon transition $\ket{-1} \leftrightarrow \ket{1}$. By introducing this detuning, $\delta$, (assuming $\delta<<\Omega$), the master equation obtained has the form of Eq.~(\ref{MasterEq}) and~(\ref{Rmatrix}) where $\varphi$ is replaced by $\varphi_{t}=\varphi_{0}+\delta t$. Hence, a sufficiently small value of $\delta$ determines the required adiabatic evolution.
Under this condition the operator $G$ assumes the form
\begin{equation}\label{eq:ruleG}
         G = \frac{\dot{\varphi}_{t}}{2} \begin{pmatrix}
                    \alpha &0&\beta\\
            0&0&0\\
                            \beta &0&-\alpha
                 \end{pmatrix}\text{,}
\end{equation}
where $\alpha=\frac{1}{\cosh 2 r}$ and $\beta=-\frac{\sinh 2 r}{\cosh 2 r}$. We show that, when $\dot{\varphi}$ is small enough,
the state $\ket{-1}\equiv\ket{\psi_{DF}}$ is adiabatically decoupled from its orthogonal subspace and a cyclic evolution in
$\varphi$ results in a geometric phase acquired by $\ket{\psi_{DF}}$ depending only, in this case, on the amount of squeezing $r$.
Note however that, since the steering process is essentially incoherent, any phase information acquired by a superposition of
$\psi_{DF}$ and a state belonging to the orthogonal subspace is inevitably lost, as the latter is subject to decoherence. The only
way to retain such information is to consider an auxiliary level $\ket{a}$, unaffected by the noise, playing the role of a
reference state for an interferometric measurement. For simplicity assume that $\ket{a}$ is unaffected by the environment during
the whole evolution, and, hence, is time independent. As a consequence, the action of the unitary transformation $O$ on $\ket{a}$
is trivial, and equation~(\ref{eq:newmastereq}) remains essentially unchanged.

The whole information about the geometrical phase and the
coherence retained by the system during its evolution is then
recorded in the phase and amplitude of the density matrix term
$\rho_{-a}=\bra{-1}\tilde{\rho}\ket{a}$, whose evolution is
described by the following set of coupled differential equations
\begin{eqnarray*}\label{eq:sys4rhoa1}
         \dot{\rho}_{-a} &=& -i \bra{a} G\rho \ket{\tilde{1}} = i
\left( \alpha\rho_{a+} -\beta \rho_{-a} \right)\frac{\dot{\varphi}}{2} \\
\dot{\rho}_{+a} &=& -\frac{\tilde{\Gamma}}{2}\rho_{+a} -i
\bra{\tilde{1}} G\rho \ket{a}
         = - \frac{1}{2}(\tilde{\Gamma} +i \alpha \dot{\varphi})\rho_{+a}
-i\beta\rho_{-a}\frac{\dot{\varphi}}{2}  \text{.}
\end{eqnarray*}
where $\rho_{+a}=\bra{1}\tilde{\rho}\ket{a}$. Assume that
initially the excited states $\ket{1}$ and $\ket{0}$ of the system
are not populated, hence $\rho_{+a}(0)=0$ and the coherence
$\rho_{-a}$ evolves as
\begin{eqnarray*}
         &&\rho_{-a}(t) =\rho_{-a}(0)\cdot\\ &&\frac{1}{\left( \lambda_{-} -
\lambda{+}
\right)}
         \left[ \left( \lambda_{-}+i\alpha\dot{\varphi} \right)e^{\lambda_{+}t}
-
         \left( \lambda_{+}+i\alpha\dot{\varphi} \right)e^{\lambda_{-}t}
\right]
         \text{,}
\end{eqnarray*}
where $  \lambda_{\pm} = -\frac{\tilde{\Gamma}}{4} \mp \frac{1}{2}
         \sqrt{\frac{\tilde{\Gamma}^2}{4} + i\alpha\tilde{\Gamma}\dot{\varphi}
- \dot{\varphi}^2  } \text{.} $

In the limit $\Gamma \gg \dot{\varphi}$ we obtain for the
coherence $\rho_{-a}$
\begin{equation}
         \rho_{-a}(t) = \rho_{-a}(0)\left( \left( 1-\epsilon \right)
         e^{i\frac{\alpha}{2}\dot{\varphi}t
         -\epsilon \tilde{\Gamma}t}+
         \epsilon e^{ -i\frac{\alpha}{2}\dot{\varphi}t -
\frac{\tilde{\Gamma}}{2}\left( 1-2\epsilon\right)
         t} \right) \text{,}
\end{equation}
where $\epsilon\simeq\frac{\beta^2}{2}\left(
\frac{\dot{\varphi}}{\tilde{\Gamma}} \right)^2$.
We are interested in a cyclic evolution, corresponding to
$T=\frac{2\pi}{\dot{\varphi}}$. By retaining only the leading
terms in $\frac{\dot{\varphi}}{\Gamma}$ the total evolution at
time $T$ is given by
\begin{equation}
         \rho_{-a} (T) = \rho_{-a}(0)e^{i\pi
\alpha
         -\alpha\beta^2\pi\frac{\dot{\varphi}}{\Gamma} }\text{,}
\end{equation}
\begin{figure}[htb]
         \begin{center}
                 \includegraphics[width=\columnwidth]{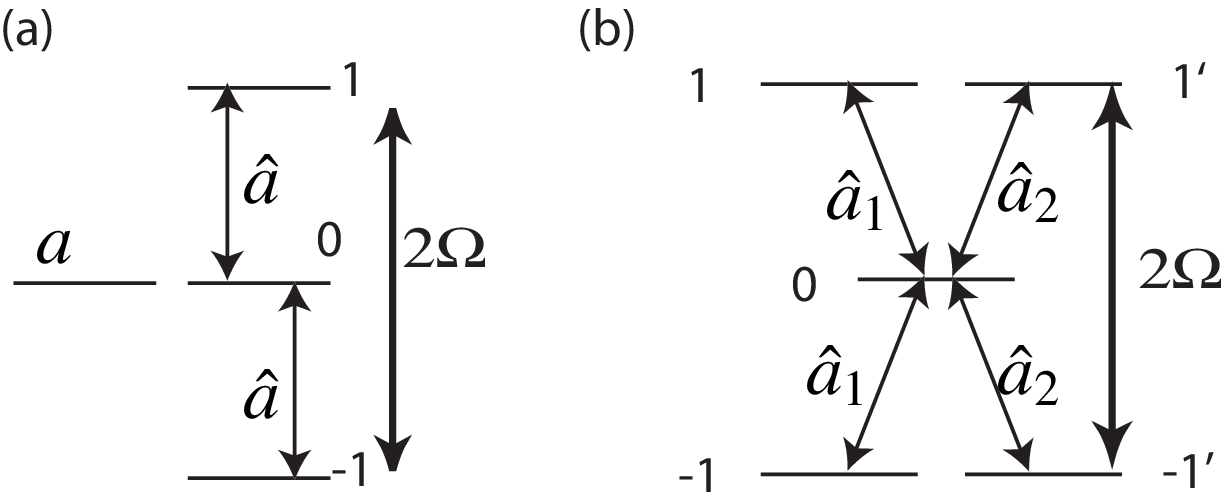}
         \end{center}
         \caption{(a) Schematic representation of the four system considered.
The energy gap between states $\ket{-1}$ and
         $\ket{0}$ and between  $\ket{0}$ and
         $\ket{1}$ is $\Omega$. The transitions between these level are coupled
to the modes $\hat{a}(\omega)$ of the reservoir.
         The reference state $\ket{a}$ is decoupled from the reservoir.\\
         (b) Five-level system, transitions $1\leftrightarrow 0$ and
$0\leftrightarrow -1$ are
         coupled to modes $\hat{a}_1(\omega)$ and $1'\leftrightarrow 0$ and
$0\leftrightarrow -1'$ are coupled to modes $\hat{a}_2(\omega)$ of the
reservoir.}
         \label{fig:sys4l}
\end{figure}
where we have substituted $\tilde{\Gamma}$ with $\Gamma/\alpha$. Finally, going back to the original frame by means of
$O^\dag(T)$), the corresponding coherence $\rho_{\psi a}(t)=\bra{\psi_{DF}}\rho(t)\ket{a}$ is given by:
\begin{equation}\label{eq:cohe}
         \rho_{\psi a} (T) = \rho_{\psi a}(0)e^{-i\pi
(1-\alpha)
         -\alpha\beta^2\pi\frac{\dot{\varphi}}{\Gamma} }\text{.}
\end{equation}
For example a state initially prepared in
$\ket{\tilde{\psi}(0)}=\frac{1}{\sqrt{2}}\left(\ket{\psi_{DF}(0)}
+ \ket{a} \right)$, after closing the loop, evolves into
\begin{equation}
         \rho (T) = \left( 1 -
         \alpha\beta^2\pi\frac{\dot{\varphi}}{\Gamma} \right)
         \ket{\tilde{\psi}(T))}\bra{\tilde{\psi}(T)}
         + \alpha\beta^2\pi\frac{\dot{\varphi}}{\tilde{\Gamma}}
\ket{\psi_{DF}}\bra{\psi_{DF}}
         \text{,}
\end{equation}
where $\ket{\tilde{\psi}(T)} = \frac{1}{\sqrt{2}}\left( e^{-i\pi
(1-\alpha)}\ket{\psi_{DF}(0)} + \ket{a} \right)$.

It is clear from this expression, that in the limit
$\xi=\dot{\varphi}/{\Gamma}\ll 1$ the dominant contribution to the
time evolution is just a phase factor $e^{i\phi}$, with
$\phi=-\pi(1-\alpha)$. This proves that in the "adiabatic"
approximation, the system preserves its coherence. In fact,
according to equation~(\ref{eq:cohe}), the amplitude damping of
$\rho_{\psi a}$ occurs only when we take into account the first
order contribution in $\xi$, which shows an exponential decay rate
of the order of $\beta^2 \frac{\dot{\varphi}}{\Gamma}$. This
proves that for small $\dot{\varphi}$ the system admits an
adiabatic limit, in which the subspace $H_{DF}(t)$ spanned by
$\ket{\psi_{DF}(t)}$ and $\ket{a}$ is adiabatically decoupled from
its orthogonal subspace $H_{\bot}(t)$. For this reason,
$H_{DF}(t)$ is decoupled from the effects of the decoherence,
which only affect states lying in its orthogonal subspace.

Within this approximation, then, a state prepared in the space
$H_{DF}(0)$ is adiabatically transported rigidly inside the
evolving subspace $H_{DF}(t)$. As a result of this adiabatic
steering, when the system is brought back to its initial
configuration, the coherence $\rho_{\psi a}$ acquires a phase
$\phi=-\pi(1-\alpha)$. This phase can be interpreted as the
geometric phase accumulated by the state $\ket{\psi_{DF}(t)}$. By
using the canonical formula for the Berry phase, it easy to see
that the geometric phase of $\ket{\psi_{DF}(t)}$ is given by
\begin{eqnarray*}
\label{eq:geomphase}
\chi_g&=&i\oint \bra{\psi_{DF}}d\ket{\psi_{DF}}=\\
&=&i\int_0^{2\pi}
\bra{\psi_{DF}}\frac{d}{d\varphi}\ket{\psi_{DF}}d\varphi=-\pi(1-\alpha)=\phi\text{.}
\end{eqnarray*}
As expected the value of $\phi$ depends only on the squeezing,
and vanishes as the squeezing tends to zero. Moreover, notice that
the phase $\phi$ is purely geometrical, i.e. there is no
dynamical contribution arising from an existing Hamiltonian,
since, in absence of any steering process, the states inside
$H_{DF}$ have a trivial dynamics. This makes the measurement of
this phase a relatively easy task. Usual procedures to measure
geometric phases make use of suitably designed techniques to
eliminate dynamical phase contributions, such as
spin-echo~\cite{JonesVEC99} 
or parallel transport
conditions~\cite{AharonovA87}. In this setup, the geometric phase is the
only contribution to the phase accumulated by $\ket{\psi_{DF}}$,
and hence, it is straightforward to measure by a suitable
interferometric setup.

A simple scheme to measure the geometric phase obtained by
such a steering process can be realized with a simple variation of
our system. Let us consider the five-level atomic system shown in
picture~\ref{fig:sys4l}(b). It essentially consists of two replicas
of the three-level system discussed above, with the level
$\ket{0}$ in common. The important ingredient is that transitions
$\ket{0}\leftrightarrow \ket{1}$ and $\ket{-1}\leftrightarrow
\ket{0}$ are coupled with modes of the reservoir which are different from
those coupled to the transitions $\ket{0}\leftrightarrow \ket{1'}$
and $\ket{-1'}\leftrightarrow \ket{0}$. A simple way to achieve
this, is to choose, for example, polarisation selective
transitions, say, left-circularly polarised modes for the former
transitions and right-circularly polarised for the latter ones.
The complete Hamiltonian of such system is:
\begin{eqnarray}\label{Hamiltonian2}
         H &=& H_S +\sum_{i=1,2}\int
\hat{a}^{\dagger}_i(\omega)\hat{a}_i(\omega)d\omega\\
         &&+ \sum_{i=1,2}\int \left[ g_i(\omega)S_i^{\dagger}\hat{a_i}(\omega)
+ H.c.
         \right] d\omega \text{,}\nonumber
\end{eqnarray}
where $H_S
=\Omega\sum_{k=-1}^1k\left(\ket{k}\bra{k}+\ket{k'}\bra{k'}\right)$,
and $S_1 =\ket{-1}\bra{0} + \ket{0}\bra{1}$ and $S_2
=\ket{-1'}\bra{0} + \ket{0}\bra{1'}$, and $\hat{a_i}(\omega)$ is
the annihilation operator of the mode with the energy $\omega$ and
polarization $i\in\{1,2\}$. Assume broadband squeezed vacuum
states for the set of modes $\hat{a}_1(\omega)$ and modes
$\hat{a}_2(\omega)$ with different squeezing parameters
$\eta_1=r_1e^{i\varphi_1}$ and $\eta_2=r_2e^{i\varphi_2}$:
\begin{equation}
\label{ }
 \ket{\text{vac}(\eta_1,\eta_2)}_{sq} = \hat{K_1}(\eta_1)\hat{K_2}(\eta_2)\vac
\text{,}
\end{equation}
 where $\hat{K_i}(\eta_i)$ are the analogous of the
operator~(\ref{eq:squeezer}) acting on the set of modes $\hat{a}_i$.
 Under the same assumptions which lead to equation~(\ref{MasterEq}) we obtain
the master equation:
\begin{equation}
\label{MasterEq2} \frac{d\rho}{dt}
=-\sum_i\frac{\Gamma_i}{2}\{R_i^\dag R_i \rho+\rho R_i^\dag R_i-2
R_i \rho R_i^\dag\}
\end{equation}
where $\Gamma_i = 2\pi | g_i(\Omega )|^2$, and $R_i(\eta_i)=S_i \cosh r_i+e^{i\varphi_i}S_i^\dag \sinh r_i$. This system admits a
two-dimensional decoherent-free subspace, spanned by states $\ket{\psi_1}$ and $\ket{\psi_2}$ whose definition is the analogous of
state $\ket{\psi_{DF}}$ of equation~(\ref{eq:DFstate}). We assume again time dependent squeezing parameters $\eta_i^t$, and again
we examine the time dependence of the system in a rotating frame, i.e. a frame where the state $\ket{\psi_i(t)}$ appear
stationary. This leads to the following master equation for the five-level system in the rotating frame:
\begin{equation}\label{eq:newmaster}
         \frac{d\tilde{\rho}}{dt} = -\sum_i\frac{\tilde{\Gamma}_i}{2}\left(
\tilde{R}_i^{\dagger}\tilde{R}_i\tilde{\rho}
         +\tilde{\rho} \tilde{R}_i^{\dagger}\tilde{R}_i -
2\tilde{R}_i\tilde{\rho} \tilde{R}_i^{\dagger} \right)
-i\sum_i[G_i,\tilde{\rho}]\text{,}
\end{equation}
where $G_i=i dO/d\eta_i O^\dag \dot{\eta}_i$, $O(t)$ being the unitary transformation producing the change of frame. Assume again,
for simplicity, that the parameters $r_1$ and $r_2$ are kept constant and that $\varphi_1=\varphi_2=\varphi$. Under this
assumption, the master equation can be exactly solved. The solution is analogous to the one obtained for system previously
analyzed. Suppose that the system is initially prepared in a coherent superposition of state $\ket{\psi_1(\eta_1^0)}$ and
$\ket{\psi_2(\eta_2^0)}$, for example:
$\ket{\psi(0)}=\frac{1}{\sqrt{2}}\left(\ket{\psi_1(\eta_1^0)}+\ket{\psi_2(\eta_2^0)}\right)$. At a later time one has
\begin{equation}
         \rho_{\psi_1\psi_2}(t) = \frac{1}{2}\exp \left\{-i(\alpha_2
-\alpha_1)\frac{\dot{\varphi}}{2}
         - \left(\frac{\beta_1^2}{2 \tilde{\Gamma_1}}+\frac{\beta_2^{'2}}{2
\tilde{\Gamma_2}}\right)\dot{\varphi}^2\right\} t
\text{,}
\end{equation}
with $\alpha_i=\frac{1}{\cosh 2 r_i}$ and $\beta_i=-\frac{\sinh 2
r_i}{\cosh 2 r_i}$. When the parameter $\varphi$ closes a loop, at
$t=T=2\pi/\dot{\varphi}$, the coherence has gained a phase
\begin{equation}
         \chi = \pi i \left( \alpha_2 - \alpha_1 \right)=\phi_2-\phi_1
\text{,}
\end{equation}
which is the difference between the geometric phases $\phi_i=\pi(1-\alpha_i)$ acquired by the states $\ket{\psi_i}$, respectively.
As in the previous scheme, the visibility is reduced by a factor which is linear in the ``adiabatic parameters''
$\dot{\varphi}/\Gamma_i$, which guarantees the existence of the adiabatic limit. The advantage of this modified scheme is that the
value of the geometric phases can be readily measured from the polarisation of the light emitted when the system relaxes. Infact,
if the value of the squeezing parameters $r_i$ is suddenly switched to zero, the states $\ket{\psi_i}$ are no longer decoherece
free, and decay to a superposition of the ground states $\ket{-1}$ and $\ket{-1'}$. This dissipation process is accompanied by two
photon emissions into the reservoir. Due to the structure of the interaction~(\ref{Hamiltonian2}) with the reservoir, the photon
emitted due to the transitions $\ket{1}\to\ket{0}$ and $\ket{1'}\to\ket{0}$, is polarised according to the geometric phase
accumulated between $\ket{\psi_1}$ and $\ket{\psi_2}$. For example, if $\hat{a}_1(\omega)$ and $\hat{a}_2(\omega)$ are right and
left circularly polarised modes, respectively, the first dissipation process produces the linearly polarised photon:
\begin{equation}
\label{polarised}
\ket{\psi_1}+e^{i\left(\varphi_1-\varphi_2\right)}\ket{\psi_2} \to
\ket{R}+e^{i\left(\varphi_1-\varphi_2\right)}\ket{L}\text{.}
\end{equation}
The detection of the polarisation of the emitted photon makes possible a direct measurement of of the geometric phase.

We have presented a scheme to generate a geometric phase via a completely incoherent control procedure. This scheme is conceptually different from the usual coherent adiabatic control. The latter is realized through a smooth evolution of suitable Hamiltonians, whereas here, the adiabatic steering is the effect of an externally controlled environment. The phase generated is purely geometrical, and, therefore, experimentally detectable without resorting to techniques for the elimination of dynamical contributions. Due to its very nature, this scheme is immune from unwanted environmental effects. Moreover, like any geometric effects, it presents an inherent degree of robustness against uncertainties in the control parameters.
\section*{Acknowledgments}
This work was supported in part by the EU under grant
IST - TOPQIP, "Topological Quantum Information Processing" (Contract IST-2001-39215). A.C. acknowledges support from Marie Curie RTN project CONQUEST. V.V. acknowledges also support from EPSRC and the British Council in Austria.


\begin{thebibliography}{10}

\bibitem{PalmaSE96}
G.~M. Palma, K.-A. Suominen, and A.~K. Ekert.
\newblock Proc. Roy. Soc. Lond., A {\bf 452} 567, (1996);
L.-M. Duan and G.-C. Guo.
\newblock Phys. Rev. Lett., {\bf 79} 1953, (1997);
P.~Zanardi and M.~Rasetti.
\newblock Phys. Rev. Lett., {\bf 79} 3306, (1997);
D.~A. Lidar, I.~L. Chuang, and K.~B. Whaley.
\newblock Phys. Rev. Lett., {\bf 81} 2594, (1998);
A.~Barenco et al. \newblock {\em SIAM J. Comput.}, {\bf 26} 1541, (1997).

\bibitem{EkertPBK89}
A.~K. Ekert, G.~M. Palma, S.~Barnett, and P.~L. Knight.
\newblock Phys. Rev. A, {\bf 39} 6026, (1989);

\bibitem{AgarwalP90}
G.~S. Agarwal and R.~R. Puri.
\newblock Phys. Rev. A, {\bf 41} 3782, (1990);

\bibitem{PoyatosCZ96}
J.~F. Poyatos, J.~I. Cirac, and P.~Zoller.
\newblock Phys. Rev. Lett., {\bf 77} 4728, (1996);

\bibitem{MyattKTSKIMW00}
C.~J. Myatt et al.
\newblock Nature, {\bf 403} 269, (2000);

\bibitem{KuzmichMP97}
A.~Kuzmich, K.~M{\o}lmer, and E.~S. Polzik.
\newblock Phys. Rev. Lett., {\bf 79} 4782, (1997);

\bibitem{CarvalhoMMD01}
A.~R.~R. Carvalho, P.~Milman, R.~L. {de Matos Filho}, and L.~Davidovich.
\newblock Phys. Rev. Lett., {\bf 86} 4988, (2001);

\bibitem{JonesVEC99}
J.A. Jones, V.~Vedral, A.~Ekert, and G.~Castagnoli.
\newblock  Nature, {\bf 403} 869, (2000);
A.~Ekert, et al.
\newblock J. Mod. Opt., {\bf 47} 2501, (2000).

\bibitem{AharonovA87}
Y.~Aharonov and J.~Anandan.
\newblock Phys. Rev. Lett., {\bf 58} 1593, (1987).

\end{thebibliography}
\end{document}